\documentclass[reprint,aps,prx,amsmath,amssymb,nobibnotes,citeautoscript,floatfix,longbibliography,raggedbottom,balancelastpage]{revtex4-2} 
\usepackage[usenames,dvipsnames]{color}
\usepackage{graphicx,xfrac,microtype}
\usepackage[bookmarks=false,colorlinks]{hyperref}
\hypersetup{hypertexnames=false,linkcolor=RubineRed,citecolor=Plum,filecolor=Mulberry,urlcolor=RoyalBlue}

\newcommand{\supf}{\textcolor{RoyalBlue}{Supplementary Fig.}\,}

\newcommand{\RN}[1]{%
  \textup{\uppercase\expandafter{\romannumeral#1}}
}

\begin{document}

\title{Thermodynamic descriptors to predict oxide formation in aqueous solutions}

\author{Lauren N.\ Walters}
\thanks{Equal author contributions.}
\affiliation{Department of Materials Science and Engineering, Northwestern University, Illinois 60208, USA}

\author{Emily L.\ Wang}
\thanks{Equal author contributions.}
\affiliation{Department of Materials Science and Engineering, Northwestern University, Illinois 60208, USA}

\author{James M.\ Rondinelli}
\email{jrondinelli@northwestern.edu}
\affiliation{Department of Materials Science and Engineering, Northwestern University, Illinois 60208, USA}

\date{\today}

\begin{abstract}
We formulate the maximum driving force (MDF) parameter as a descriptor to capture the thermodynamic stability of aqueous surface scale creation over a range of environmental conditions.
We use formation free energies, $\Delta_f G$s, sourced from high-throughput density functional theory (DFT) calculations and experimental databases to compute the maximum driving force for a wide variety of materials, including simple oxides, intermetallics, and alloys of varying compositions.
We show how to use the MDF to describe trends in aqueous corrosion of nickel thin films determined from  experimental linear-sweep-voltometry data.
We also show how to account for subsurface oxidation behavior using depth-dependent effective chemical potentials. 
We anticipate this approach will increase overall understanding of oxide formation on chemically complex multielement alloys, where competing oxide phases can form during transient aqueous corrosion. 
\end{abstract}

\maketitle

\section{Introduction}

Quantitative comparisons of predictive scale formation in corrosion-resistance alloy design remains a grand challenge in materials and corrosion sciences.
A general approach is to rely on free energies of formation of the bulk oxide, $\Delta_f G$, and its location on the convex hull (composition-energy diagram)
to predict which oxide is most likely to form. This approach omits key aspects of the oxidation problem at the nanoscale.
Initial ($\leq$10$^{-7}$\,s) and kinetically-controlled film growth can be described by decoupled kinetic models, such as percolation and diffusion models \cite{Percolation_Model, Experimental_kinetic_1, Experimental_kinetic_2}, however, these rely often on experimentally-derived parameters that are challenging to extract.
On the other hand, thermodynamic phase diagrams have demonstrated success as predictive tools for understanding scale growth \cite{Liang_Nickel_MainPourbaix, Liang_NiThinFilms_Pourbaix, General_Poubaix_1, General_Pourbaix_2, General_Pourbaix_3, General_Pourbaix_4}. Predominance diagrams, including Pourbaix \cite{M_Pourbaix, Thompson_MultielementPourbaix}  and stability diagrams \cite{StabilityDiagrams1, StabilityDiagrams2, StabilityDiagrams3}, compare how the environment described by pH, potential, concentration, and temperature change the stability regions of ions and solids. 
\autoref{fgr:mdf-calculation}a shows the stable phases in the potential--pH space 
for the Cu-H$_2$O system.
The phase diagram-based models, however, are computationally intensive to calculate for multiple-element systems, \textit{e.g.}, a binary or multi-principal elemental alloy.
They also do not provide a direct way to compare oxide-phase stabilities among different alloy systems; they only convey the size of the stability fields. 
%

Here we use the concept of thermodynamic driving forces, \textit{i.e.}, chemical potential differences, as a way to quantitatively describe aqueous stable (hydr)oxide formation during electrochemical oxidation.
High thermodynamic driving forces are likely to indicate which initial phases will appear, and may also describe phase evolution and the final equilibrium microstructure that forms. 
%
For example, steel-process design (quenching or annealing) to achieve  diffusionless transformations for the creation of martensitic transformation-induced-plasticity  steels may be informed by driving forces, determined from chemical potential lines with respect to temperature and composition \cite{steel_transformations_1, steel_transformations_2}. 
%
Recently, the synthesis science community has used chemical potential mapping to select the most successful precursors for target-phase synthesis \cite{myBiMOQ, GreenRust, StabilityDiagrams_Riman} and mapping of the initial reaction steps verified by \textit{in situ} characterization support the use of thermodyamic driving forces \cite{Ceder_Thermo_Kinetics, Ceder_chem_potent}.

In situations where early solid oxide growth may be followed by aqueous dissolution, a high driving force for scale formation would likely produce a solid-liquid equilibrium with an oxide phase at the boundary separating the alloy from the aqueous environment.
Herein, we formulate the maximum driving force (MDF) descriptor to characterize   
solid phase oxide or hydroxide formation in an aqueous metal-H$_2$O system.
We calculate it from the chemical potential difference between the solid element or (hydr)oxide and the most stable aqueous ion over a range of pH and potentials.
We implement a workflow that leverages existing computational materials science frameworks and experimental thermodynamic databases to calculate the MDF for systems with $n\leq5$ elements in equilibrium with H$_2$O.
Furthermore, we introduce a second metric to account for oxide growth parametrically through effective oxygen chemical potential changes ($\mu^\mathit{eff}_\mathrm{O}$) without explicitly treating the kinetics of nucleation and growth.
We show how to use these descriptors to interpret experimental film growth studies of the aqueous nickel (hydr)oxide system.
Next, we examine the MDF trends for transition-metal and main-group elements, focusing on correlations between the MDF and enthalpy of formation.
%
Finally, we propose the MDF parameter can be used in a predictive manner to assess oxide phase evolution from different metals, thereby informing alloy composition design for selective oxidation.

\section{Model and Methods}

\subsection{Thermodynamic Descriptor Formulation}

Materials in aqueous environments may be stable or react with water or other dissolved ions to form aqueous dissolution products or solids, most commonly in the form of oxides or hydroxides. 
Corrosion resistant materials often rely on the creation of a stable, solid (hydr)oxide, termed a scale, to limit any soluble release of ions or larger particulate formation.
Native (hydr)oxide M$_x$O$_y$H$_z$ formation is
described using a generalized redox reaction of water with metal $\mathrm{M}$ as 
\begin{equation*}
    X\mathrm{M}+Y\mathrm{H}_2\mathrm{O} \rightarrow \mathrm{M}_x\mathrm{O}_y\mathrm{H}_z + (2Y-Z)e^-+(2Y-Z)\mathrm{H}^+ 
    \label{hydroxide-rxn}
\end{equation*}
where $X$, $Y$, and $Z$ provide stoichiometric  mass balance.
To predict if a solid phase forms, we calculate the  chemical potentials ($\mu)$ for each species.
The chemical potential of solid elements in their most stable phase at standard state ($T=298.15$\,K, $P=1$\,atm) is $\mu=\Delta_f G= 0$, where $\Delta_f G$ is the bulk Gibbs free energy of formation. 
The chemical potential ($\Delta \mu_{\rm solid}$) for the formation of a solid phase, \emph{e.g.},  M$_x$O$_y$H$_z$, from the reaction of water with its respective metal is 
\begin{align}\label{eq:hydroxide-chem-pot} 
    \Delta \mu_{\mathrm{M}_x\mathrm{O}_y\mathrm{H}_z} 
    &= \bigl(  \Delta _f G_{\mathrm{M}_x\mathrm{O}_y\mathrm{H}_z} -(2Y-Z)\cdot FU \nonumber\\ 
    &- (2Y-Z) \cdot RT\text{ln}(10) \cdot \mathrm{pH} -X\cdot \Delta _f G_{\rm M} \nonumber\\ 
    &- Y\cdot \Delta _f G_{{\rm H}_2{\rm O}}\bigr)/N\,, 
\end{align}
where $R$ is the ideal gas constant, $T$ is the temperature, $F$ is Faraday's constant, and $N$ is the total number of metal elements per formula unit (here $N=X$).
Corrosion occurs through the solubilization of the solid and subsequent formation of aqueous ions $[\mathrm{M}_x\mathrm{O}_y\mathrm{H}_z]^\delta_\mathrm{(aq)}$, for which the reaction chemical potential $\Delta \mu _{\rm rxn}$ to form the aqueous ion $\Delta \mu_{\rm aq.\,ion}$ is
\begin{align}\label{eq:aq-chem-pot}
      \Delta \mu_{[\mathrm{M}_x\mathrm{O}_y\mathrm{H}_z]^\delta_\mathrm{(aq)}} 
      &= \bigl(
      \Delta _f G_{[\mathrm{M}_x\mathrm{O}_y\mathrm{H}_z]^\delta_\mathrm{(aq)}} +RT\text{ln}(\eta_I) \nonumber\\ 
      &-(2Y-Z+\delta)\cdot FU \nonumber\\
      &- (2Y-Z)\cdot RT\text{ln}(10) \cdot \mathrm{pH} \nonumber\\
      &- X\cdot \Delta _f G_{\rm M} - Y\cdot \Delta _f G_{{\rm H}_2{\rm O}} \bigr)/N\,, 
\end{align}
where $\delta$ is the oxidation state and $\eta_I$ is the solute activity (set to zero for a solid).
The first two terms of \autoref{eq:aq-chem-pot} show that the  chemical potential contributions to the aqueous ion include both its free energy of formation and some measure of  solvation captured by the solute activity.
\autoref{eq:hydroxide-chem-pot} and \autoref{eq:aq-chem-pot} may be formulated to include different reacting species known to be present in water, such as OH$^-$ (alkaline conditions), H$^+$ (acidic conditions), and common aqueous ions (\textit{e.g.}, M$^{\delta}$).

\begin{figure}
\centering
\includegraphics[width=0.99\linewidth]{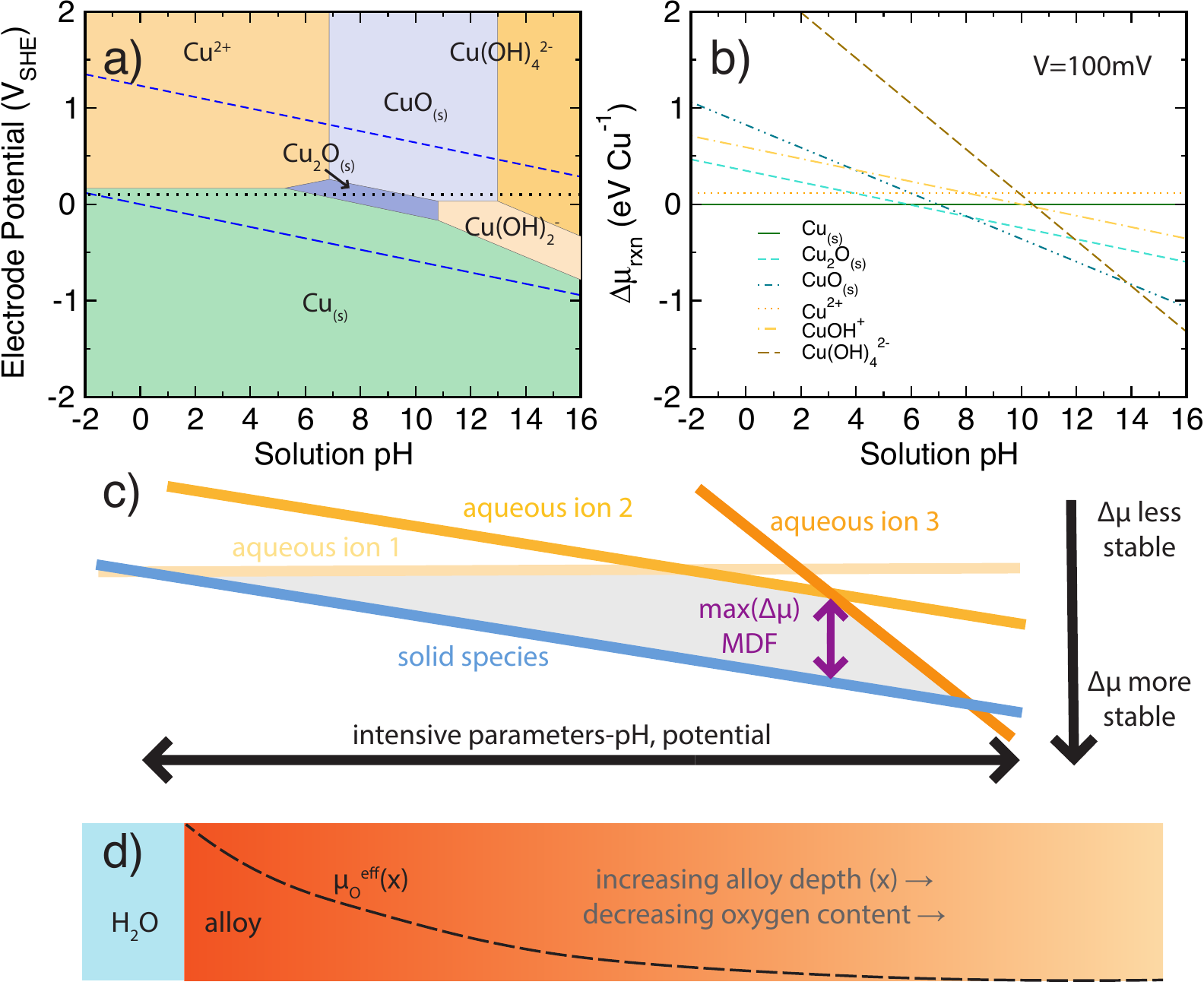}
\caption{
(a) Cu Pourbaix diagram showing the pH and potential ranges, elemental states (teal), solid oxide formation representative of passive copper layers (blues), and corrosive aqueous ions (oranges). 
(b) Driving force diagram at 100\,mV for pHs in between -2 and 16. 
(c) Schematic multidimensional driving force diagram projected into two-dimensions showing how the MDF is calculated by computing differences between driving force planes that appear here as lines.
(d) Schematic of $\mu_\mathrm{O}^\mathit{eff}$(x), where oxygen content and chemical potential decrease with increasing depth into the alloy. 
}
\label{fgr:mdf-calculation}
\end{figure}

These expressions are used commonly to construct Pourbaix diagrams.
For example, taking a constant potential contour through \autoref{fgr:mdf-calculation}a (dotted line at applied potential of 100\,mV), the stable phases appearing here are obtained by examining the 
relative chemical potentials of aqueous copper species at varying pH values and identifying those with lowest $\Delta \mu _{\rm rxn}$ (\autoref{fgr:mdf-calculation}b). 
The small chemical potential differences between the first and second most stable species in \autoref{fgr:mdf-calculation}b indicate there  are small driving forces for Cu formation at low pH values, whereas there are moderate driving forces to stabilize a solid oxide for $6\leq \mathrm{pH} \leq14$.
This missing variation in magnitude of the driving force to form  oxides in \autoref{fgr:mdf-calculation}a is what motivates us to compute the potential and pH dependent MDF.

The maximum driving force (MDF) for solid phases that would nucleate on the surface of a metal in an aqueous system is defined as  
\begin{equation}\label{eq:mdf-definition}
    \Delta \mu_{\rm max} =\max\left(\Delta \mu_{\rm solid}-\Delta \mu_{\rm aq.{ }ion}\right) \equiv \mathrm{MDF}\,,
\end{equation}
where $\Delta \mu_{\rm max}\le0$ is the optimized driving force, $\Delta \mu_{\rm solid}$ is the reaction chemical potential for solid phase formation, \textit{e.g.}, an alloy or (hydr-)oxide, obtained from \autoref{eq:hydroxide-chem-pot}, and $\mu_{\rm aq.\,ion}$ is the reaction chemical potential for the most stable aqueous ion(s) at a given set of environmental conditions found in \autoref{eq:aq-chem-pot}.
Although  \autoref{eq:mdf-definition} does not have any interfacial or kinetic terms explicitly included, we make the ansatz that rapid formation of the solid is expected whenever there are large driving forces, because the liquid provides sufficiently fast transport of ions.

Computing the MDF requires: 
(\textit{i}) identifying the most stable solid-aqueous ion pair and then (\textit{ii}) then calculating the chemical potential difference at the specific electrochemical conditions, \textit{i.e.}, pH, potential, concentration, for which the greatest difference occurs. 
Section\,\ref{sec:methods} provides additional details on free energy sources and numerical evaluation details. 
A schematic of the MDF shown in \autoref{fgr:mdf-calculation}c, illustrates that the chemical potential lines enclose a range of conditions (shaded area) that stabilize an oxide more than aqueous ions. The purple arrow indicates the difference in solid phase and aqueous ion chemical potential at maximizing conditions.
Details for extrapolating the MDF to multiple elements are provided in the SI. 

The environmental conditions are defined first prior to calculating the MDF. 
The constraints are required because the calculated stability of solid scales and aqueous ion species evolves through the environmental condition space, including pH, potential and concentration ranges (see the Supporting Information, SI).
Therefore, unless otherwise noted, we defined the default set of conditions as:  $2.5\leq\mathrm{pH}\leq12.5$, -500\,mV\,$\leq$\,V$_{\rm SHE}\,\leq$\,750\,mV, $\eta_I$\,=\,10$^{-6}$, $T=25^\circ$C, and $P=1$\,atm, here termed the standard corrosion limit window.

Initial solids most likely to form at the surface are the most thermodynamically stable species based on the overall highest driving force within a chemical potential window. 
Although $\Delta \mu_{\rm max}$ assesses oxide formation at the surface of a material from a difference in stability between aqueous and solid species, subsurface oxidation exclusively occurs as a solid-state transformation. 
Below this initial layer, time-dependent oxidized solid formation is reflective of the compositionally-constrained subsurface environment.
Therefore, we use a model that is unconstrained at the surface, with fast reaction rates due to mobile water and ions, and without knowledge of the solid bulk composition \cite{Ceder_Thermo_Kinetics}.
We assume the subsurface system to be isothermal and isobaric, and an open system to oxygen but closed to other elements. (See Section II.C for more on this assumption.)

Following Ref.\ \onlinecite{Ceder_Thermo_Kinetics,*Ceder_chem_potent},
the grand canonical free energy 
for a given (solid) product, \textit{e.g.}, oxide or hydroxide, below the surface is
$ \bar{\phi}^{\rm eff}_{\rm solid} = ({\Delta _f G_\mathrm{solid} -N_\mathrm{O}\mu^{\rm eff}_\mathrm{O}(x)})(\sum N_{\rm metal})^{-1}$, 
where $\Delta _f G_{\rm solid}$ is the free energy of formation of the product, $N_\mathrm{O}$ and $N_{\rm metal}$ is the number of oxygen and metal atoms per formula unit, respectively, and $\mu^{\rm eff}_\mathrm{O}(x)$ is the depth-dependent effective oxygen chemical potential.
Here, we parametrically decrease $\mu^{\rm eff}_\mathrm{O}$ (or another ion such as hydrogen,  $\mu^{\rm eff}_\mathrm{H}$) with scale depth to reflect declining oxygen content below a solid's surface, for which the true depth dependence is a consequence of diffusion during active corrosion and/or changes in oxygen solubility from alloy processing. In other words, as is shown in \autoref{fgr:mdf-calculation}d, the oxygen chemical potential decreases accordingly with the reduction in local oxygen composition.
The chemical potential of the product will then decrease according to oxygen composition. 

\subsection{Methods\label{sec:methods}}

\subsubsection{Data Sourcing}
For single-element--H$_2$O systems, two different experimental sources for solid free energies of formation were used: experimental energies from Pourbaix's \emph{Atlas of Electrochemical Equilibria in Aqueous Solutions} \cite{M_Pourbaix} and accurate DFT energies simulated with the hybrid function HSE06, reported in multiple studies by L.F. Huang \cite{Liang_NiThinFilms_Pourbaix, Liang_Nickel_MainPourbaix}. Experimental ion free energies of formation were acquired by combining data from Pourbaix and Materials Project \cite{MP_DFT, general_MP} to obtain a comprehensive list of ions.

\subsubsection{Constructing Driving Force Diagrams}

Driving force diagrams were created by calculating the chemical potential planes for each of the individual products or product combinations in the system. In multi-element systems, there can be multiple products corresponding to one plane in the driving force diagram, which leads to exponentially increasing possible combinations as further elements are added. We modified the reduction scheme \cite{convex_hull_red} in the \texttt{PourbaixDiagram} module available in PyMatgen \cite{pymatgen, MP_PourbaixMod_2} by separately reducing all-solid and all-ion combinations. This ensures we retain all necessary combinations for an accurate calculation of the MDF. Chemical potential planes were then computed for only the compositionally possible \cite{Thompson_MultielementPourbaix} combinations from the reduction step. 

\subsubsection{Calculating the Maximum Driving Force}

The product combinations and their chemical potentials are binned into three groups: solid products (often oxides and hydroxides), aqueous ion products, and misfit products (relevant for multi-element systems, where product combinations involved in the system do not fit within the aforementioned criteria for the two categories. An example of this product is a mixed-solid and ion product. The most stable solid, corrosion, and misfit product and their respective chemical potential values are stored within a pH-potential phase space with a sample grid density of 125,000 points.
We note that the considered solids are filtered to ensure stability in water and code efficiency (see SI). The most stable solids' chemical potential and the most stable aqueous ions' chemical potential are then subtracted to yield driving force values across the phase space.
In extended multi-element systems, driving forces for solid products at pH-potentials where misfit products are most stable overall are excluded. 

\section{Results and Discussion}

\subsection{Bulk and {pH/V} Dependent {Ni} Film Growth}

\begin{figure}
\centering
\includegraphics[width=0.95\linewidth]{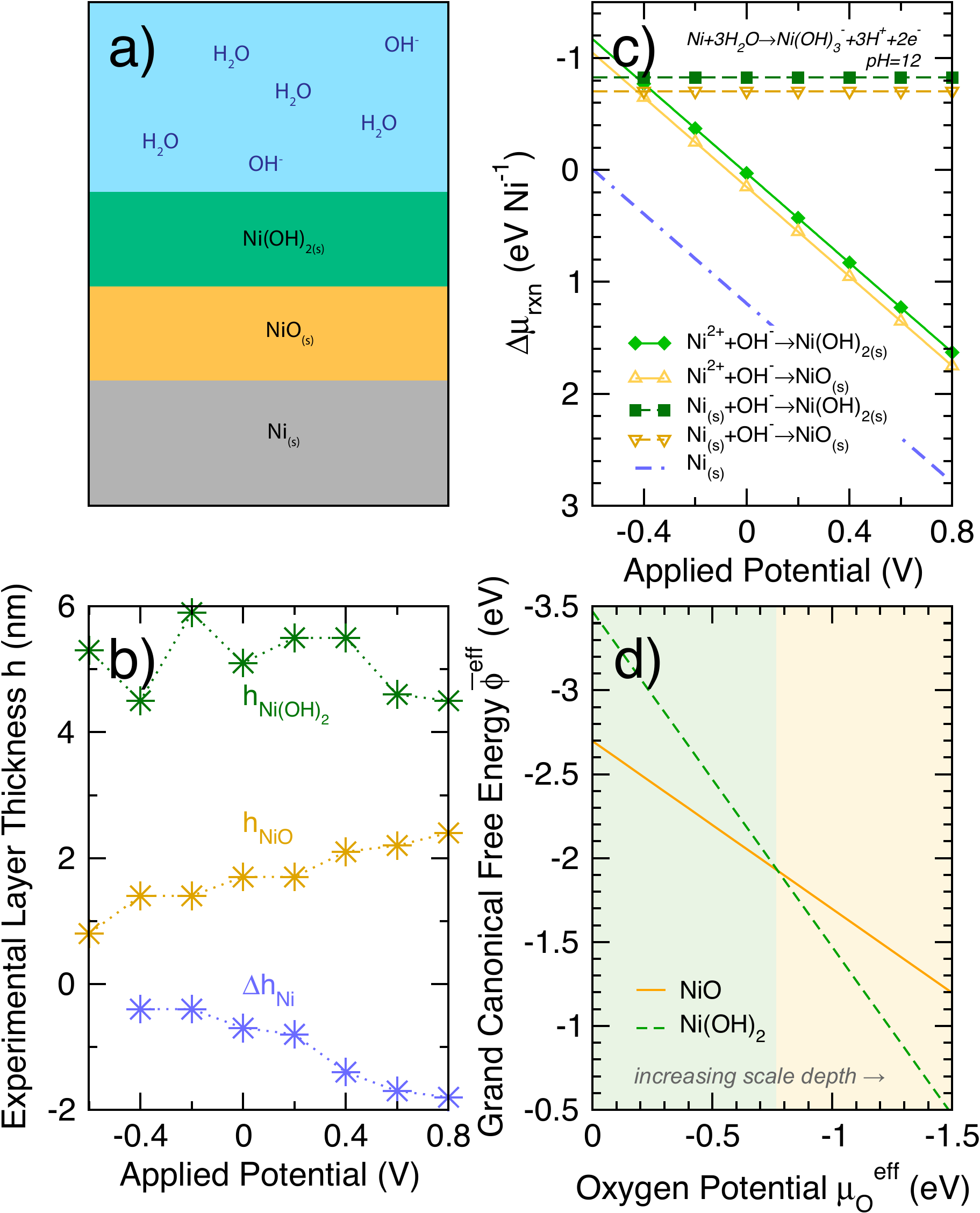}
\caption{
(a) Schematic of the Ni thin film system constructed by Huang \textit{et al.} \cite{Liang_NiThinFilms_Pourbaix}, where Ni films are subjected to specific pHs and then characterized at distinct applied potentials. At most pHs, Ni(OH)$\rm _{2(s)}$ forms at the surface and it then transforms to NiO below the surface Ni(OH)$\rm _{2(s)}$ layer.
(b) Ni thin film depth ($h$) dependence at pH\,=\,12 measured in Huang \textit{et al.} \cite{Liang_NiThinFilms_Pourbaix}. 
(c) The driving forces of thin films as a function of applied potential,  -600\,mV\,$\leq$\,V$_{\rm SHE}$\,$\leq$\,800\,mV, calculated in comparison to the most stable ion Ni(OH)$_3^-$. Ni(OH)$\rm _{2(s)}$ exhibits the maximal (most negative) driving forces to form at all voltages, and results in the deepest film depth of any solid.
(d) NiO$\rm _{(s)}$ and NiOH$\rm _{2(s)}$ depth-dependent stability, calculated from $\bar{\phi}^{\rm eff}$ as $\mu^{\rm eff}_\mathrm{O}$ is varied. At a relatively low oxygen chemical potential of $\mu^{\rm eff}_\mathrm{O}=-0.7$\,eV with limited oxygen, NiO$_{\rm(s)}$ becomes the most stable solid. Note the chemical potential axes are flipped.
}
\label{fgr:ni}
\end{figure}

We now use the MDF to understand scale formation of nickel (hydr-)oxides thin films.
Generally, Ni(OH)$_{\rm 2(s)}$ initially grows at the surface and then NiO$_{\rm (s)}$ evolves beneath it at moderate pHs and potentials (approximately 7\,$\leq$\,pH\,$\leq$\,15, -0.5\,V\,$\leq$\,V$_{\rm SHE}\,\leq$\,1\,V) (\autoref{fgr:ni}a) \cite{Liang_NiThinFilms_Pourbaix, Ni_3, Ni_4}. Primary surface NiO$_{\rm (s)}$ initiates at lower pHs between 5 and 7. Huang \textit{et al.}  characterized this pH-dependent and depth-dependent nickel scale formation by reporting film identity and thickness at pH\,=\,4.9 and 12.0 for distinct potentials in the range of -600\,mV\,$\leq$\,V$_{\rm SHE}$\,$\leq$\,800\,mV \cite{Liang_NiThinFilms_Pourbaix}.
\autoref{fgr:ni}b shows their experimental Ni thin film depths, $\Delta h_{\mathrm{Ni}_{(\mathrm{s})}}$, at pH\,=\,12 decrease nearly 2\,nm with increasing oxidation potential from about -0.6\,V to 0.8\,V. 
Simultaneously, the subsurface NiO$\rm _{(s)}$ film depth increases in magnitude from only 0.8\,nm to 2.4\,nm. The thickness $h\rm_{Ni(OH)_{2(s)}}$ starts at $\approx5$\,nm, demonstrating fast initial growth, but plateaus during further oxidation potential change.
This and other past literature utilized model Pourbaix diagrams and interpreted this  phase evolution using $\Delta _f G$ values, but could not fully capture the  pH-dependent and depth-dependent scale formation \cite{Liang_NiThinFilms_Pourbaix, Liang_Nickel_MainPourbaix, Scully_NiFe_Pourbaix}.

\autoref{fgr:ni}c presents the driving forces found through the subtraction of the chemical potential of Ni(OH)$_3^-$ from the formation of solid oxides and element Ni.
Consistent with previous reports \cite{Liang_Nickel_MainPourbaix,Liang_NiThinFilms_Pourbaix}, we choose $\Delta_f G$s calculated from DFT HSE06 calculations to better model the aqueous electrochemical behavior of nickel.
Surface energy corrections are not included here, but could though inclusion of surface electronic energy 
\cite{Liang_NiThinFilms_Pourbaix}. 
We find the most stable solid scales to be Ni(OH)$\rm _{2(s)}$ and NiO$\rm _{(s)}$,  consistent with previous thermodynamic predictions \cite{Liang_Nickel_MainPourbaix, Liang_NiThinFilms_Pourbaix, M_Pourbaix}. At alkaline pHs, OH$^-$ (rather than H$_2$O) and Ni$^{2+}$/Ni (potential-dependent) exhibit the highest (most negative) driving forces, fostering Ni(OH)$\rm _{2(s)}$ scale formation. Ni(OH)$\rm _{2(s)}$ exhibits higher unconstrained (bulk) driving forces  than NiO formation at all potentials, consistent with its initial significant growth on the Ni film ($h\rm _{Ni(OH)_{2(s)}} \approx5$\,nm). The dot-dash line reveals that elemental Ni is increasingly unstable and prone to oxidation as the potential increases, through either solubilization into Ni(OH)$_3^-$ or scale formation, which agrees with the disappearance of Ni film depth in \autoref{fgr:ni}b.
The initial and stable nickel hydroxide surface film is a product of the large chemical potential difference between Ni(OH)$_{\rm 2(s)}$ and the most stable ion, Ni(OH)$_3^-$. 
The less negative driving forces for surface nickel oxide formation appear in \autoref{fgr:ni}d, where NiO forms as the most thermodynamically stable product under subsurface composition constraints (constrained $\mu_\mathrm{O}^{\rm eff}$).
Finally, we show for thin films grown at pH\,=\,4.9, driving forces shown in the SI 
favor NiO surface growth, consistent with experimental characterization.
Therefore, the MDF and $ \bar{\phi}^{\rm eff}_{\rm solid}$ successfully describe the external and internal oxidation of the Ni film.

\subsection{Mapping Elemental Corrosion Trends} 
\begin{figure}
\centering
\includegraphics[width=0.99\linewidth]{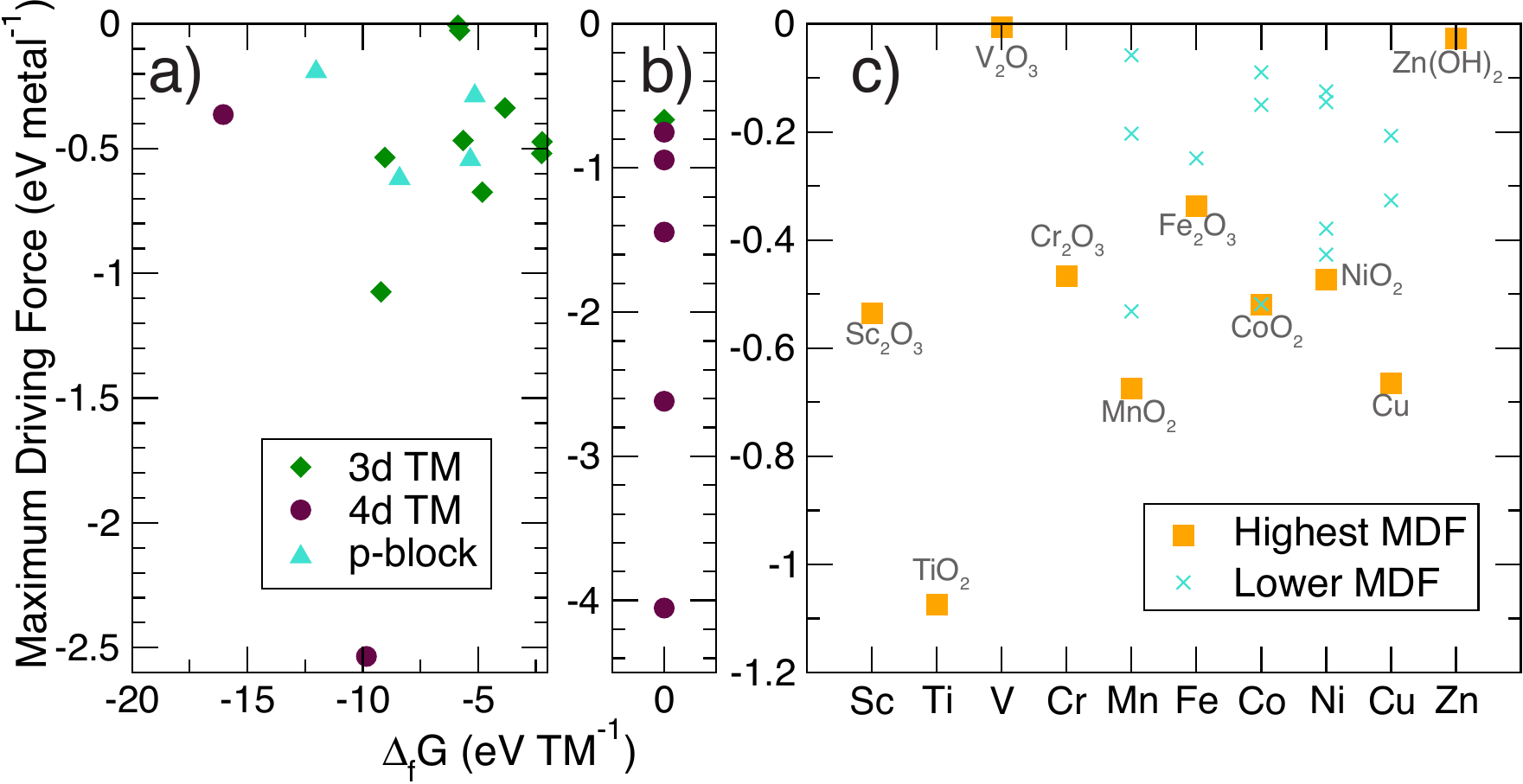}
\caption{
(a) Calculated MDF versus free energy of formation for 3\textit{d} and 4\textit{d} transition metals (TM), and select \textit{p}-block elements (Al, Ga, Sn, In) whose solid product is an oxide, hydroxide, or oxyhydroxide.
(b) The calculated MDF for 3\textit{d} and 4\textit{d} elements whose MDFs are calculated from their elemental states, \emph{i.e.}, $\Delta _f G=0$.
(c) Calculated MDF for each of the transition metals, including the most stable (orange, $\blacksquare$) and any other stable (teal, $\times$) solid products.
All calculations are performed for the standard corrosion window with thermodynamic data sourced from experimental $\Delta_f G$ values \cite{M_Pourbaix}.
}
\label{fgr:periodic-table-mdf}
\end{figure}

We now explore MDF trends for elements by calculating the MDF for
$3d$, $4d$, and select main group metals, using $\Delta_f G$ values sourced from Pourbaix's Atlas \cite{M_Pourbaix}.
\autoref{fgr:periodic-table-mdf}a-b shows the stable solid (element, oxide, hydroxide, or oxyhydroxide) with the highest MDF in the standard corrosion window. 
The MDFs are plotted against Gibbs free energy of formation for the solid, which has been utilized as a ``rule of thumb'' guide for assessing phase stability \cite{Wolverton_freeenergy_spinels}.
First, we note that MDF and $\Delta_f G$ show little correlation.
The elements group within one of three categories according to the two energy descriptors:  (\textit{i}) $ \Delta_f G\approx$\,0 and MDF\,$<$\,0, demonstrating little thermodynamic drive for general oxidation despite the MDF indicating solid stability in water,  (\textit{ii}) $ \Delta_f G<$\,0 and MDF\,$\approx$\,0, occurring for a solid that readily might form outside of water, but  has an unstable or metastable scale within an aqueous environment, or (\textit{iii}) $ \Delta_f G<$\,0 and MDF\,$<$\,0 where there is high thermodynamic drive within and outside aqueous conditions to generate a solid oxidation product. The MDF, as it represents a driving force for scale formation over aqueous ion formation (corrosion), is a unique descriptor which captures behavior missing in $\Delta_f G$.

\autoref{fgr:periodic-table-mdf}a-b demonstrates also the variability in scale stability and corrosion behavior that exists within the periodic table. The 4\textit{d} elements exhibit the largest MDFs, such as Ru and Ag, whose elemental metals have MDFs of -4.05\,eV and -2.61\,eV, respectively. The 4\textit{d} row also contains two elements, Nb and Cd, that have an MDF close to 0, indicating metastable oxide or hydroxide scales that may be more susceptible to other effects (\textit{e.g.} kinetic effects) within the standard corrosion window. This leads to a discrepancy between the MDFs of Nb and Cd and their reported solid oxide phase formation in water, which we attribute to our data set's inclusion of the Nb(OH)$_{\rm 5(aq)}$ and Cd(OH)$_{\rm 2(aq)}$ ions, whose very negative $\Delta _f G$ values limit known solid stability. This inconsistency highlights the need for high-fidelity energetic data and the importance of data sources, particularly for high throughput studies that may employ the MDF for predictive purposes.

Nearly all the 3\textit{d} transition metals exhibit MDFs ranging from -0.25 to -0.75\,eV/metal. All 3\textit{d} TMs but copper find their largest driving forces with a (hydr)oxide product. Moreover, V$_2$O$_3$ is the only oxide scale predicted from the 3\textit{d} series, again demonstrating the need for accurate sourcing of $\Delta _f G$ values, particularly hydroxide values from DFT, which exhibit greater uncertainity (\supf1-3).
The main group metals also show moderate MDFs. In(OH)$_3$ exhibits the largest MDF at -0.62\,eV/In and the smallest MDF is for hydrated Al$_2$O$_3$ with -0.19\,eV/Al.
Finally, as expected many noble metals, such as Ag and Cd, have a high driving force to resist oxidation ($\geq1$\,eV), despite $\Delta _f G =0$ (\autoref{fgr:periodic-table-mdf}b).

To further explore variation in the 3\textit{d} row, we plot all solids with $\mathrm{MDF}\leq0$ for each element in \autoref{fgr:periodic-table-mdf}c.  Data shown in squares indicate the species with the highest MDF per element. The crosses represent any solids with MDFs less than the most stable species. 
We find that TiO$_2$ has the largest driving force of all 3\textit{d} oxidized phases explored. %
Most other 3\textit{d} transition metals exhibit MDFs ranging from approximately -0.4 to -0.7\,eV per transition metal. Cu is the only 3\textit{d} metal whose elemental form is most stable within the standard corrosion window with respect to oxidized products. Furthermore, between Mn and Cu, we find multiple oxides, hydroxides, and oxyhydroxides provide sizeable MDFs. This data suggests that Mn, Fe, Co, Ni, and Cu are more likely prone to multiphase competition during electrochemical oxidation.
Further use of $\bar{\phi}^{\rm eff}_\mathrm{solid}$ applied to these metals may provide insight into subsurface compound formation, particularly for materials with multiple metastable phases.

\subsection{Additional Considerations}
The MDF allows for relatively easy to interpret and low resource-intensive predictions for systems of multiple different element systems or varying compositions.
This is particularly relevant to advanced alloy systems, such as multi-principal element alloys (MPEAs) with 5 or more elements, for which recent research focuses on multi-component interactions, but is resource demanding \cite{HEA_Corrosion_1, HEA_Corrosion_2}. It has been shown that the superior corrosion resistance of some MPEAs can be linked to mixed-metal oxides formed via higher driving forces than relevant binary oxides \cite{alloys_superiorpassivation}. 
To that end, effects of alloying elements may be compared with the MDF to understand any increased or suppressed driving force, in addition to any changes to scale  composition or structure. 
Furthermore, the MDF can provide increased understanding of alloy composition by 
comparing $\Delta _f G$ values  with respect to alloy composition. 
%
%
In practice, thermodynamic competition between solids of similar (M)DFs may also provide insight into complex or phase-separated scales.
Last, we note this approach requires either calculated or experimentally determined accurate energies for new alloying system or compositions, and this becomes the bottleneck to using the MDF broadly.

Reaction kinetics can play a dominant role in the formation of new materials.
We propose  two kinetically driven aspects of the system for which the MDF 
may approximate: (\textit{i}) reaction rate of the initial surface species formation, and (\textit{ii}) sub-surface, depth-dependent scale formation. As in Ref.\ \onlinecite{Ceder_Thermo_Kinetics}, we argue that the initial species to form will be that with the highest driving force. 
Because a reaction rate is often approximated as the ratio of the thermodynamic driving force to a generalized resistance, the spatial or temporal effects hindering the transformation must be defined. 
At the solid-aqueous boundary, there should be little resistance for the transformation. 
Thus, the thermodynamic driving force prescribes the dynamics of the system. 
Solid-solid transformations, described next, would require defining resistance terms to evaluate whether a transformation would proceed; examples, include anion and cation transport, defect density, interfacial reaction rates, and thickness, among others.
Furthermore, state-of-the-art protection attributes, which can tune resistance terms  such as sacrificial cathodic protection in the presence of Cl$^-$, may be incorporated my modifying the  potential to account for additional ions, leaving of any less noble metallic species, and additional reactions to form complexing soluble ions or multiple-anion solid surface phases within \autoref{eq:hydroxide-chem-pot}-\ref{eq:aq-chem-pot} \cite{state_of_art_coatings}.

Internal (subsurface) oxidation within aqueous thin film growth are limited by ion mobility, \textit{e.g.}, diffusion barriers of redox agents such as H$^+$ and OH$^-$ below the aqueous-solid interface. 
Oxidized products that occur subsurface have not been described by widely used, easily-calculable parameters. 
Spatially-dependent chemical potentials can describe concentration gradients, as  used in phase field models \cite{phasefield1, phasefield2,phasefield3_voorhees}, for example 
$\mu_\mathrm{O}^\mathit{eff}(x)$ and $ \mu_\mathrm{H}^\mathit{eff}(x)$, to 
examine internal/external oxidation and serve as a resistance proxy for diffusional barriers of O and H from the aqueous medium. By graphing the driving forces with respect to constrained chemical potentials, we can examine species evolution both from reaction of elements/alloys and from the transformation of initial surface products.
Future experimental work to characterize depth-dependent oxygen content or film formation through use of x-ray photoelectron spectroscopy, time of flight secondary ion mass spectrometry, or reflectometry measurements may be used to assess the bounds on $\mu_\mathrm{O}^\mathit{eff}(x)$.

\section{Conclusions}

In conclusion, we formulated quantitative descriptors for solid phase formation in aquas environments using free energies of formations for solids and ions. 
We shows the maximum driving force parameter enables one to compare the tendency for solid phase formation in aqueous electrochemical environments among different materials systems, which is difficult to do from pH--potential phase diagrams.
The MDF and depth dependent descriptors presented herein are versatile and easily-calculable based on available thermodynamic databases, and are therefore ideal for implementation in high-throughput workflows.
We propose using the MDF to guide the selection for alloying elements and composition to control corrosion behavior, understanding depth-dependent scale growth, and devising  solvothermal synthesis methodologies.\\

\section*{Data and Code Availability}
Data and code related to algorithms that implement \autoref{eq:hydroxide-chem-pot}-\ref{eq:mdf-definition} and are used in the calculation of multi-element  and compositionally-dependent MDFs and driving-force diagrams are deposited on \href{https://github.com/MTD-group/MDF}{Github}.

\begin{acknowledgments}
L.N.W. sponsored by the Department of Navy, Office of Naval Research (ONR), under ONR Award number N000014-16-12280. E.L.W. was supported by the National Science Foundation’s MRSEC program (DMR-1720139) at the Materials Research Center of Northwestern University as an Undergraduate Research Intern. J.M.R. was sponsored under ONR Award number N00014-20-1-2368. The United States Government has a royalty-free license throughout the world in all copyrightable material contained herein.
\end{acknowledgments}

\bibliography{References}
\end{document}